\documentclass{moriond}

\usepackage{amsmath}
\usepackage{multicol}
\usepackage{amssymb}
\usepackage{moresize}
\usepackage{subfig}

\def\be{\begin{equation}}
\def\ee{\end{equation}}
\def\bea{\begin{eqnarray}}
\def\eea{\end{eqnarray}}

\begin{document}
\vspace*{4cm}
\title{Development of an unbiased cosmic shear estimator measured on galaxy images}

\author{\textbf{Enya Van den Abeele}, Pierre Astier, Anna Niemiec}
\address{LPNHE, CNRS-IN2P3 and Universités Paris 6 \& 7}

\maketitle\abstracts{
Since cosmic shear was first observed in 2000, it has become a key cosmological probe and promises to deliver exquisite dark energy constraints. However, shear is inferred from coherent distortions of galaxy shapes, and the relation between galaxy ellipticities and gravitational shear is a serious potential source of bias. To address this, we are developing a shear estimation method that makes no assumption on galaxy shapes, in order to avoid the shortcomings of a simulation-based shear calibration. Our method relies on the estimation of second moments on the image, and the evaluation of how second moments respond to a shear applied to the coordinate system, without altering the image itself, at variance with the Metacalibration method. We also evaluate analytically the noise bias due to the non-linearity of the estimator, and confront it with the bias derived from noisy image simulations, which allows a fast and precise noise bias correction.}

\section{Introduction}
Data from future-generation surveys like Vera C. Rubin Observatory (LSST) will pave the way for precision cosmology. With an image density of $\sim$37 galaxies/arcmin², and a total of 20 billion galaxies at the end of the full survey data releases, LSST is the first ground-based telescope designed for weak lensing. Because of its sensitivity to both matter and expansion, weak lensing (and more specifically cosmic shear) has become a powerful tool to understand dark energy, being the most sensitive probe to constrain its equation of state parameters $w_0$ and $w_a$ according to LSST's year-10 forecast\cite{lsst}. But despite the unprecedented image's galaxy density provided by this telescope, the shear measurement remains something complex and associated with biases. These biases are usually parameterized as follows\cite{huterer2006} : 
\begin{equation}
    g^{obs} = (1+m)g^{true} + c
    \label{shear_bias_eq}
\end{equation}
with $g^{obs}$ the observed shear of a source, $g^{true}$ its true shear, $m$ the multiplicative bias and $c$ the additive bias. There are many possible sources that can lead to a multiplicative bias, including a poorly calibrated shape estimator or the noise present in the image. Because a bias on the shear measurement introduces a bias on the cosmological parameters estimation, a limit to the multiplicative factor value - compatible with the LSST statistic - must be set : according to Cropper \textit{et al} 2013\cite{cropper2013}, the limit we need to achieve on the bias is on the order of $10^{-3}$. With this in mind, we propose a new approach for shape measurement and calibration independent of the galaxy profile described in section \ref{estimation_method}, and an analytical solution to the noise bias described in section \ref{noise_bias}.

\section{Shear estimation method and technical aspects}
\label{estimation_method}
The first thing that we need in order to perform a shear estimation is to measure the shape of the sources affected by cosmic shear, in our case galaxies. Whatever the estimator, the basic principle relies on the fact that the \textit{second moments} of a galaxy image are altered by shear. In the continuous limit, the second moments are expressed as follows : 
\begin{equation}
\begin{split}
    M &= \int (X-X_0)(X-X_0)^TW(X-X_0)I(X-X_0) \,d^2X\ \\
    &= \int YY^TW(Y)[I_0 \circledast \psi](Y) \,d^2Y\
\end{split}
    \label{moments1}
\end{equation}
with $X$ the image coordinates and $X_0$ the object position, $W$ the weight function (introduced to optimize the signal-to-noise ratio) and $I$ the image (resulting from the convolution between the above atmosphere galaxy image $I_0$ and the Point Spread Function (PSF) $\psi$).\\
This formula gives a 2x2 matrix thanks to which we can define the ellipticity estimator $\textbf{e}$, that describes the shape of the galaxy :
\begin{equation}
    \textbf{e} = 
\begin{pmatrix}
e_1 \\
e_2
\end{pmatrix} 
=
\begin{pmatrix}
M_{xx} - M_{yy} \\
2M_{xy}
\end{pmatrix} 
\label{ellip}
\end{equation}
Defining $\psi_{-}(X) \equiv  \psi(-X)$, equation \ref{moments1} can be rewritten as follows (taking $X$ as $X-X_0$) :
\begin{equation}
    M = \int [(XX^TW(X))\circledast \psi_{-}](X)I_0(X) \,d^2X\
    \label{moments3}
\end{equation}
where we used Parseval's identity and the convolution theorem. The challenge in designing the estimator lies in measuring the shear on galaxies with unknown shapes, which can introduce bias. Even assuming galaxies are randomly oriented, the ellipticity derived from second moments isn't enough due to the PSF contribution, requiring a calibration. Moreover, trying to accurately simulate galaxy profiles is a really non-trivial exercise, but distorting images is more controlled. The goal is to establish a shear-sensitive estimator reliant on minimal assumptions, particularly about galaxy profiles.\\
The introduction of a shear will transform the image plane and coordinates (from $X$ to $SX$ to first order) before the transformation by the PSF. Starting from this, we can introduce an artificial shear to the original galaxy image $I_0$ and then reconvolve the distorted galaxy image with the PSF to recover the initial configuration. This is essentially what is done in \textit{Metacalibration}\cite{metacal}. Another solution can be to use equation \ref{moments3} to apply the shear not to $I_0$, but to the other terms of the estimator (i.e. : $F(X)=([XX^TW]$ $\circledast$ $\psi)(X)$) :
\begin{equation}
    M(S) = \int F(S^{-1}X)I_0(X) \,d^2X\
    \label{moments5}
\end{equation}
where $S$ is the 2x2 shear matrix, with det($S$) = 1. Then, we can recover the original image $I(X)$ by dividing $F(Sk)$ by $\psi$ in Fourier space, which gives $G(S,X)$ in real space :
\begin{equation}
    M(S) = \int G(S,X)I(X) \,d^2X\
    \label{moments6}
\end{equation}
These distorted second moments will be used to define the derivatives. In practice, because we chose to work with a sampled PSF, we need to calculate numerical derivatives : using the $S$ matrix, we define 4 shear variations ($\pm\epsilon$ on $g_1$ and same for $g_2$) that we apply to the coordinates system to distort the pixel grid. We then interpolate the $F$ function on these new grids, which leads to 4 new images, for which we calculate the respective distorted second moments (noted $M^S_{1\pm}$ and $M^S_{2\pm}$). Because we are working on sampled images, the second moments we get from the image are the sum of the continuous moments (equation \ref{moments1}) and a \textit{pixel second moments}. As we measure the $M^S$ matrices on a distorted grid, we should subtract a \textit{distorted} pixel second moments matrix to recover the real object one. We can rewrite a new formalism for $M$ :
\begin{equation}
    M(s,\epsilon) \propto \gamma + \alpha\epsilon + \alpha'\epsilon^2 + \beta s^2 + \beta' s^4 + \delta s^2\epsilon + \delta' s^4\epsilon
    \label{corr}
\end{equation}
where $s$ is the image's pixel scale. The $\gamma + \alpha\epsilon  + \alpha'\epsilon^2$ term represents the continuous sheared second moments, $\beta s^2 + \beta' s^4$ the sampling correction, and the term including $\delta'$ the cross-effect between shear and sampling. Since the $\beta$ and $\beta'$ terms are only involved in the $M_{xx}$ and $M_{yy}$ components, they cancel each other out because of the subtractions in \textbf{e} and \textbf{R}. Thus, the only term we need to put into the correction to perform a shear estimation is the last one, $\delta'$.
Then, after this correction applied to the distorted $M^S$, the derivatives are calculated as follows :
\begin{align*}
    \frac{\partial M}{\partial g_1} &= \frac{M^S_{1+} - M^S_{1-}}{2\epsilon} &  \frac{\partial M}{\partial g_2} &= \frac{M^S_{2+} - M^S_{2-}}{2\epsilon}
\end{align*}
Thanks to these derivatives, we can define the self-calibration factor $\textbf{R}$ :
\begin{equation}
    \textbf{R} = 
\begin{pmatrix}
\frac{\partial e_1}{\partial g_1} & \frac{\partial e_2}{\partial g_1}\\
\frac{\partial e_1}{\partial g_2} &
\frac{\partial e_2}{\partial g_2}
\end{pmatrix} 
=
\begin{pmatrix}
\frac{\partial M_{xx}}{\partial g_1} - \frac{\partial M_{yy}}{\partial g_1} &
\frac{2\partial M_{xy}}{\partial g_1} \\
\frac{\partial M_{xx}}{\partial g_2} - \frac{\partial M_{yy}}{\partial g_2} &
\frac{2\partial M_{xy}}{\partial g_2} 
\end{pmatrix}
\label{Rfactor}
\end{equation}
Given $\textbf{e}$ and $\textbf{R}$, we can then define an estimator of the shear  $\langle g \rangle$ :
\begin{equation}
    \langle g \rangle = \langle\textbf{R} \rangle^{-1} \langle e \rangle
    \label{shear_moy}
\end{equation}
The advantages of this methods are multiple : all the calculations are based on second moments, rather than model fitting methods, so we don’t have to make any assumption about the galaxy profile. Moreover, the $F$ function is more extensive than the “above atmosphere” image $I_0$, it is therefore better to apply shear distortion on it, because distorting $I_0$ introduces correlated noise. It also allows to perform shear estimations on under-sampled galaxy images.

\section{First results on noise-free simulations}
\begin{figure}[h!]
    \centering
    \subfloat{\includegraphics[scale=0.28]{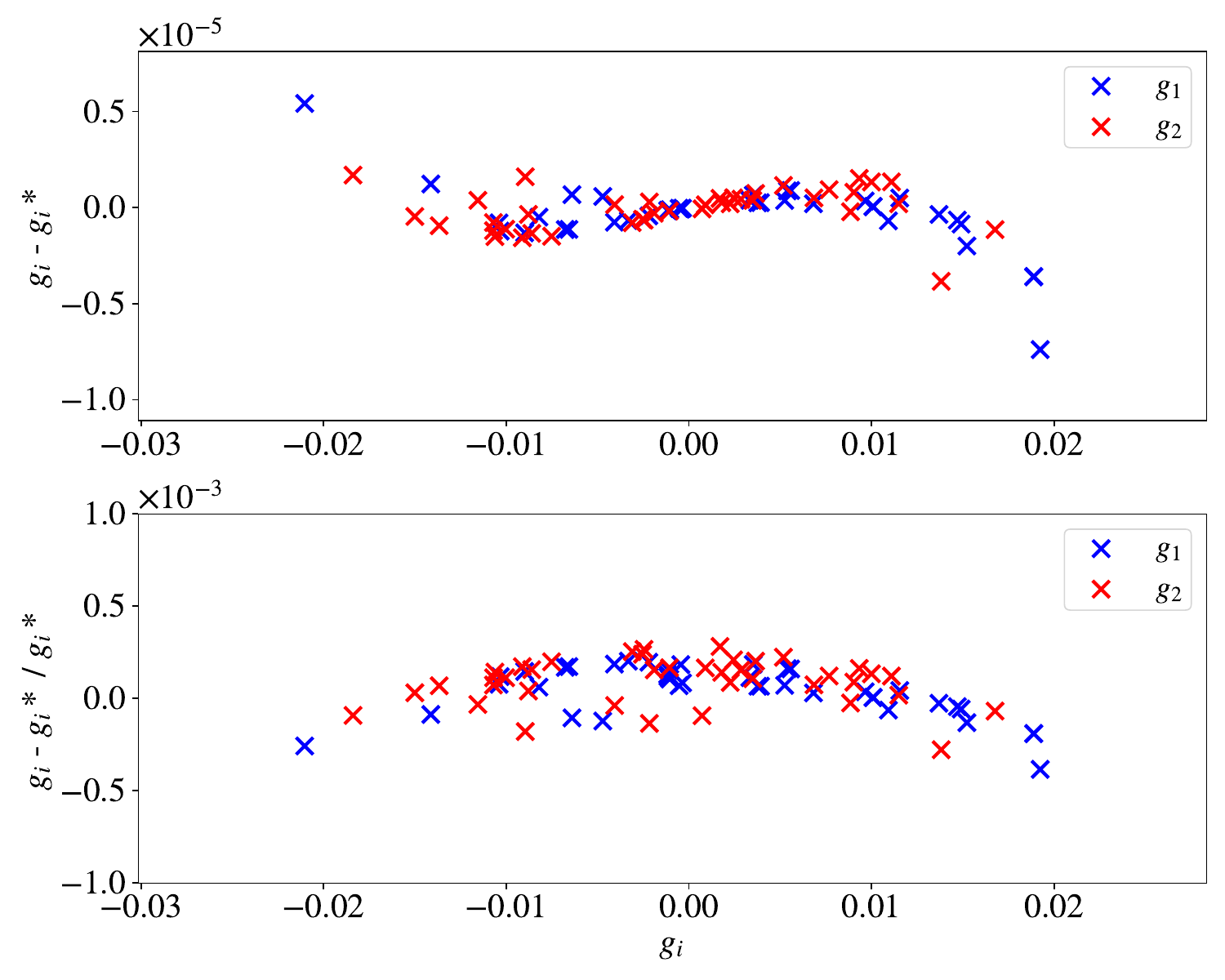}}%
    \qquad
    \subfloat{\includegraphics[scale=0.28]{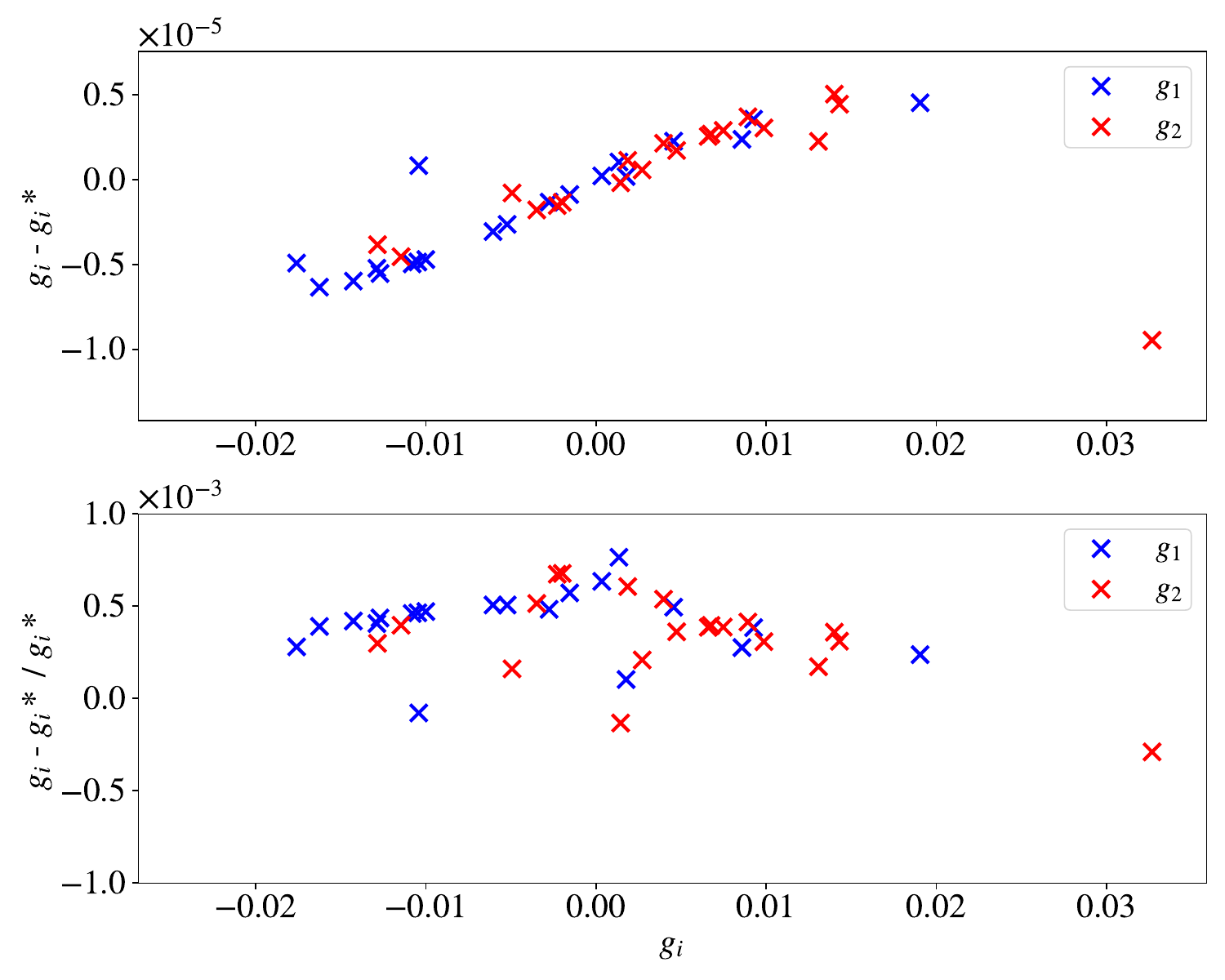}}%
    \caption{\begin{footnotesize}Absolute (top) and relative (bottom) differences between input and output shear values ($g_1$ blue and $g_2$ red). Estimation performed on elliptical galaxies (left) and COSMOS galaxy (right).\end{footnotesize}}%
    \label{shear_est_plots}%
\end{figure}

In order to test the estimator performances, we set different galaxy simulations, starting from elliptical Gaussians to more realistic profiles from the \textit{COSMOS} catalog. We also tested for different PSF profiles (Gaussians, Kolmogorov and Moffat), and we choose a Gaussian weight function that is identical for all galaxies in the same size range, whose second moments are equivalent to those of considered galaxy sample images. All the simulations were performed thanks to the $\textit{Galsim}$\cite{galsim} package. Carried out over 40 random shear values and averaged over 20 pairs of random (and opposite) intrinsic ellipticities, these estimations show very satisfying results, with a bias under our of $10^{-3}$ upper limit (see figure \ref{shear_est_plots}).

\section{Noise bias analytical correction}
\label{noise_bias}
\begin{figure}[h!]
    \centering
    \subfloat{\includegraphics[scale=0.28]{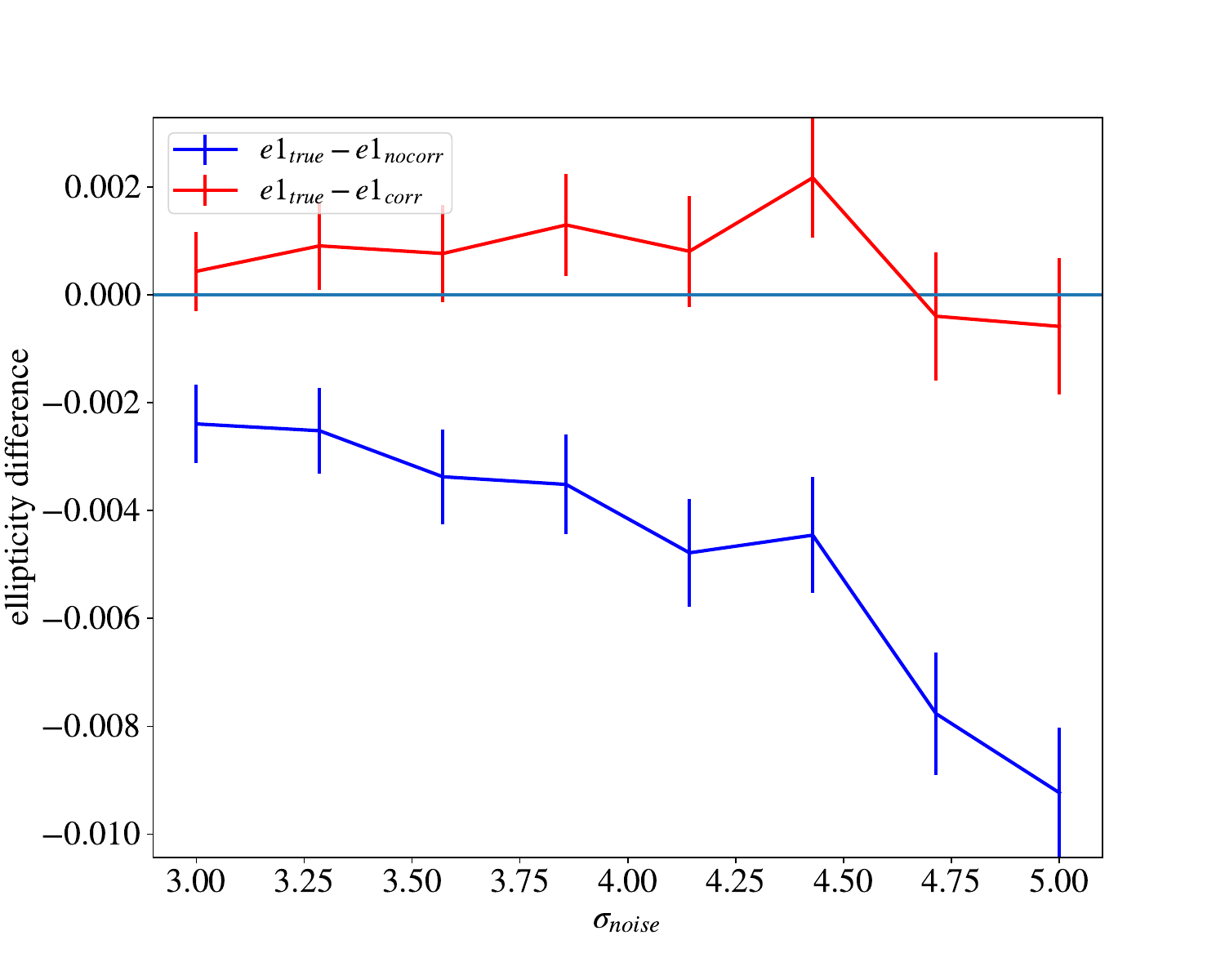}}%
    \qquad
    \subfloat{\includegraphics[scale=0.28]{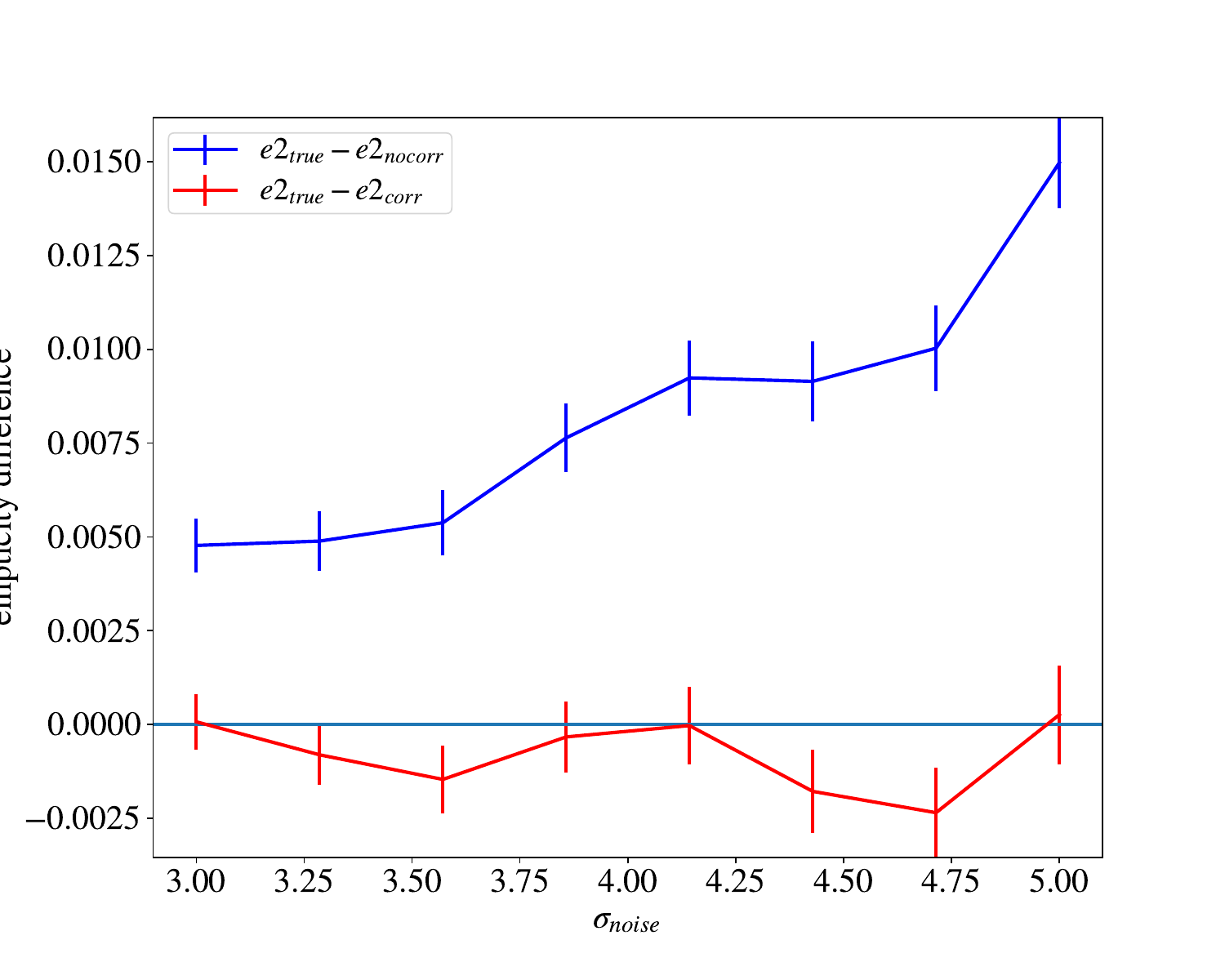}}%
    \caption{\begin{footnotesize}First (left) and second (right) parameters of ellipticity calculated from noisy second moments (blue) and corrected seconds moments using the analytical noise bias prediction (red), as a function of $\sigma_{noise}$.\end{footnotesize}}%
    \label{ellip_plots}%
\end{figure}

As we measure the position of the galaxy directly on the received - and noisy - image, it creates a bias in the second moments calculation. This position is indeed re-injected into the pixel coordinates and the weight function $W$ (see formula \ref{moments1}). The dominant term of the bias affecting the second moments can be estimated by evaluating the second derivatives of the moment with respect to the image, considering its dependency on the position $x_0$ :
\begin{small}
\begin{align}
 m_I(I+n) &= m_I(I) + \sum_k \frac{\partial m_I}{\partial I_k}n_k + \frac{\partial m_I}{\partial x_0} \delta x_0  \nonumber \\ 
&+ \frac{1}{2} \left(\sum_{kl} \frac{\partial^2m_I}{\partial I_k \partial I_l} n_k n_l + \sum_{k} \frac{\partial^2m_I}{\partial I_k \partial x_0 } n_k \delta x_0  + \frac{1}{2} \frac{\partial^2m_I}{\partial x_0^2 } \delta x_0 \delta x_0  \nonumber \right)+ ...
\end{align}
\end{small}
As we are taking the average on a set of realizations, and we assume the mean of the noise to be zero, the linear terms disappear. It is only necessary to calculate the quadratic terms, in other words $\partial^2m_I/\partial x_0^2$ and $\partial^2m_I/\partial I_k \partial x_0$. This leads to several correction terms, depending on the position variance and the size of the different components involved in $M$.
This noise bias correction applied to $M$ gives successful results on the estimation of \textbf{e}, as shown on figure \ref{ellip_plots}. The same kind of noise bias analytical formulas needs to be calculated for the distorted $M^S$, in order to correct \textbf{R}, but this is still a work in progress.
\vspace{-0.5cm}

\section{Conclusion}
To achieve precision cosmology analysis with cosmic shear thanks to future LSST data, we need to limit the multiplicative bias on shear measurement to $10^{-3}$. In this context, we developed an unbiased self-calibrated shear estimator independent of the galaxy profile without applying any distortion to the galaxy image, and giving satisfying results on basic tests. Furthermore, we calculated analytical formulas to correct the noise bias affecting the estimator, with a good correction on ellipticity and promising results on the calibration factor.

\end{document}